\documentclass[%
superscriptaddress,
 amsmath,amssymb,
 aps,
 pra,
]{revtex4-2}

\usepackage{graphicx}
\usepackage{dcolumn}
\usepackage{bm}
\usepackage[capitalise]{cleveref}
\usepackage[mathlines]{lineno}
\usepackage{soul}

\begin{document}

\preprint{APS/123-QED}

\title{Single-shot X-ray Dark-field Tomography}

\author{Adam Doherty}
\email{adam.doherty@ucl.ac.uk}
\author{Ian Buchanan}
\author{Alberto Astolfo}
\author{Savvas Savvidis}
\affiliation{Department of Medical Physics and Biomedical Engineering, University College London, London, WC1E 6BT, UK}
\author{Mattia F. M. Gerli}
\affiliation{UCL Division of Surgery and Interventional Science, Royal Free Hospital, NW3 2PF, London, UK}
\affiliation{Stem Cell and Regenerative Medicine Section, Great Ormond Street Institute of Child Health, University College London, London, WC1N 1EH, UK}
\author{Antonio Citro}
\affiliation{San Raffaele Diabetes Research Institute, IRCCS San Raffaele Scientific Institute, Milan, Italy}
\author{Alessandro Olivo}
\author{Marco Endrizzi}
\affiliation{Department of Medical Physics and Biomedical Engineering, University College London, London, WC1E 6BT, UK}

\date{17th August 2023}

\begin{abstract}
X-ray dark-field imaging creates a representation of the sample where contrast is generated by sub-resolution features within the volume under inspection. These are detected by a local measurement of the radiation field's angular distribution, and how it is affected by the interaction with matter. X-ray dark-field imaging typically requires taking multiple exposures for separating the contributions to the detected X-ray intensity arising from scattering, refraction and attenuation; a procedure often called phase retrieval. We propose an approach to retrieve an X-ray dark-field image from a single X-ray shot. We demonstrate the method using a laboratory-based, rotating anode X-ray tube system without the need for coherent radiation or a high-resolution detector. This reduces the complexity of data acquisition, enabling faster scanning and increasing dose efficiency. Moreover, our approach reduces the problem dimensionality by one, with substantial implications for data-intensive applications like tomography. The model assumes a homogeneous material, and we show this is a valid hypothesis for soft biological tissues by reconstructing dark-field tomography images from data sets containing a single shot per view. We believe our method to be broadly applicable and relevant for many X-ray dark-field imaging implementations, including fast radiography, directional dark-field and for use with pulsed X-ray sources.
\end{abstract}

\maketitle

In X-ray dark-field and phase-contrast imaging, contrast is generated by the phase changes imparted to the radiation as it traverses the sample. Sensitivity to these contrast channels requires the use of specialised setups, often implemented with synchrotron radiation that offers high flux and coherence, but also adapted to laboratory-scale equipment. We focus here on edge illumination \cite{olivo2021edge}, for its applicability with rotating anode X-ray sources  \cite{olivo2007coded}, robustness \cite{zamir2016robust}, negligible coherence condition (both spatial and temporal) \cite{munro2010source} and because it is the only approach we know of that allows for uniform sampling of the illumination across a cm-sized field of view \cite{doherty2020optimal}, which is important for single-shot imaging.

In phase-sensitive X-ray setups, the intensity reaching the detector is modulated by both attenuation and phase effects and separating these signals into attenuation, (differential) phase and dark-field images typically requires taking multiple exposures under different conditions, such as different propagation distances or positioning of optical elements or modulators \cite{langer2008quantitative, rigon2007generalized, pfeiffer2008hard, endrizzi2014hard}.  Alternatively, this can be done with a single exposure by finely sampling the illumination but comes at a cost of reducing the system resolution below that of the detector \cite{vittoria2015beam, kagias20162d, zhou2018single}. With single-shot imaging, we aim to extract the signal of a chosen contrast channel from only a single X-ray intensity projection. This necessarily requires assumption about the sample but potentially offers faster, more robust, more efficient and simpler acquisition protocols. 

We introduce a homogeneous-object approximation for a solution that removes the need for multiple X-ray projection exposures or higher pixel density for a local analysis of the wavefront. This was inspired by the vastly applied Paganin retrieval \cite{paganin2002simultaneous} that allows for phase imaging from a single X-ray intensity projection, which was also recently translated to dark-field imaging in the context of synchrotron radiation \cite{beltran2023fast}. We demonstrate our approach with a lab-based system, and we show the benefits on a data-intensive application such as tomography.

\begin{figure*}
    \centering
    \includegraphics[width=\linewidth]{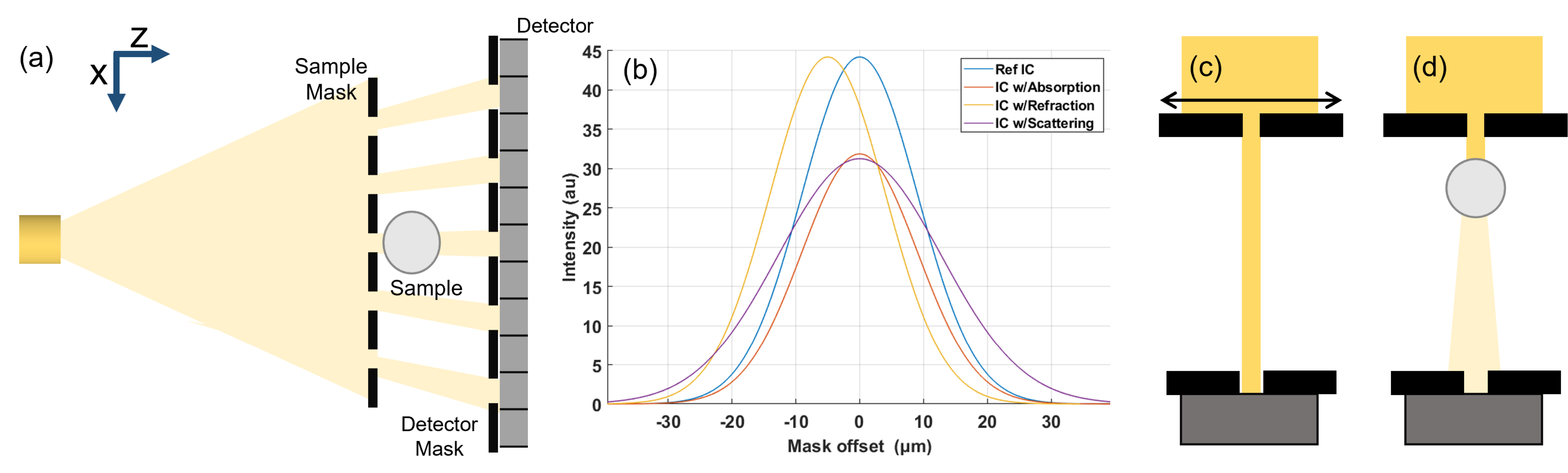}
    \caption{(a) Edge illumination system schematic, with radiation from an X-ray tube source modulated by the sample mask before reaching the sample, and then travelling towards the detector mask and detector. (b) Illumination curves, the intensity acquired by one pixel when the sample mask is translated along the x-axis through one period (see arrow in (c)), and the effect on them for purely attenuating, refracting and scattering samples. (c) - (d) Single beamlets are shown for a pictorial description of those effects, for the mask position corresponding to the peak in the illumination curve. A drop of intensity is expected at this position from both attenuation as well as scattering effects, associated with the dark-field signal.}
    \label{fig:EI_IC}
\end{figure*}

A schematic of an edge illumination system is shown in \cref{fig:EI_IC}(a). The system uses an absorbing mask that has a one-dimensional periodic structure that modulates the beam into a series of beamlets. A second mask is placed just before the detector to analyse the profiles of these beamlets, with phase sensitivity along the direction between beamlets, and each beamlet associated with one detector pixel. The experiments in this work were performed with the prototype system developed at UCL, a full description of which can be found in Havariyoun et al. \cite{havariyoun2019compact}.

Imaging with the edge illumination system is through the acquisition and processing of each pixel's illumination curve. This is the intensity measured by one pixel as the sample mask is translated in one period, and it is measured by taking a series of exposures at different offset positions between the sample mask and the detector mask along the $x$-axis. A model illumination curve is shown in \cref{fig:EI_IC}(b), for which a Gaussian function is a well-established approximation \cite{endrizzi2014hard}. The dark-field image is retrieved through fitting Gaussian functions to the data acquired for each illumination curve \cite{astolfo2016ret}, with the change in the area under the illumination curve corresponding to attenuation, the shift in centre position to refraction or differential phase, and the increase in width to scattering or dark-field contrast. In the case of tomography \cite{doherty2023edge}, the procedure needs to be repeated for each viewing angle, and thus the amount of images required is the number of sampling points along the illumination curve times the number of viewing angles, $N_{IC} \times N_{proj}$. In an edge illumination system, spatial resolution can be improved through a procedure known as dithering, where the sample is translated in small steps along the $x$-axis, subsequently exposing parts of the sample that have not been illuminated, further increasing the acquisition dimensionality to three and the number of exposures to $N_{IC} \times N_{dith} \times N_{proj}$.  The model and methodology presented here allows quantification of the dark-field signal with a stationary sample mask (i.e. no illumination curve scanning, $N_{IC} =1$), reducing the acquisition dimensionality by one for all dark-field acquisitions. The proposed method additionally does not require curve fitting and hence it also substantially speeds up an otherwise computationally intensive data processing.  

\section*{\label{sec:ss_methods}Single-Shot Dark-Field Model}

The basis for single-shot dark-field imaging with edge illumination is a convolution model of the illumination curve. The illumination curve without the sample is denoted as the reference $I_r(\bar{x})$:
\begin{equation} \label{eq:refIC}
    I_r(\bar{x}) = \frac{A}{\sqrt{2\pi\sigma_r^2}}\exp\left [ -\frac{(\bar{x}-\bar{x}_0)^2}{2\sigma_r^2} \right ]
\end{equation}
where $\bar{x}$ indicates the displacement of the sample mask along the $x$-axis, $A$ is a constant determining the amplitude of the curve, $\sigma_r^2$ is the variance, and $\bar{x}_0$ is the position of alignment between the sample and the detector apertures. Note that these parameters $[A, \bar{x}_0, \sigma_r^2]$ are all pixel-wise and vary between illumination curves across the detector. Variations in $\bar{x}_0$ can originate from mask imperfections or misalignment. Single-shot dark-field retrieval requires imaging at the same point on each illumination curve across the detector, meaning variations in $\bar{x}_0$ should be minimised but these can be reduced to well below a micron across the whole field of view \cite{doherty2020optimal, brombal2023pepi}. We will assume $\bar{x}_0$ to be constant for all pixels. Variations in $\sigma_r^2$ are less critical and can be accounted for by pixelwise correction, but we can assume this parameter to be constant without a significant performance loss. 

When a sample is in the beam, the illumination curve is denoted as the sample illumination curve $I_s(\bar{x})$ and includes three additional terms:
\begin{equation} \label{eq:sampIC}
    I_s(\bar{x}) = \frac{tA}{\sqrt{2\pi(\sigma_r^2+\sigma_o^2)}}\exp\left [ -\frac{(\bar{x}-\bar{x}_0-\Delta \bar{x}_{ref})^2}{2(\sigma_r^2+\sigma_o^2)} \right ]
\end{equation}
with $t$ quantifying the change in area from attenuation, $\Delta \bar{x}_{ref}$ quantifying the lateral shift from refraction, and $\sigma_o^2$ quantifying the broadening in angular distribution from scattering (dark-field). Calculating these three parameters on a pixel-by-pixel basis yields attenuation, refraction and dark-field contrast images. 

The effects of the attenuation and refraction signals must be isolated or excluded for retrieving $\sigma_o^2$ from a single measurement. Sensitivity to refraction is linked to the gradient of the illumination curve \cite{diemoz2015single}, and hence it is minimised at the peak and tails, which is also where sensitivity to the dark-field broadening is highest. Sensitivity to attenuation signal is highest at the peak (See \cref{fig:EI_IC}(d)). For single-shot dark-field imaging, we chose to expose at the peak of the illumination curves, where refraction effects can be neglected and sensitivity to scattering is retained, reducing the number of unknowns to two ($t$ and $\sigma_o^2$). We also note that this illumination is the most dose efficient with little X-ray intensity lost in the analyser. 

At the peak of the illumination curve, i.e. when $\bar{x}=\bar{x}_0$, the measured signal for the reference and sample illumination curves are as follows:
\begin{equation} \label{eq:refIC_peak}
    I_r(\bar{x}=\bar{x}_0) = \frac{A}{\sqrt{2\pi\sigma_r^2}},
\end{equation}
\begin{equation} \label{eq:sampIC_peak}
    I_s(\bar{x}=\bar{x}_0) = \frac{tA}{\sqrt{2\pi(\sigma_r^2+\sigma_o^2)}}\exp\left [ -\frac{\Delta \bar{x}_{ref}^2}{2(\sigma_r^2+\sigma_o^2)} \right ].
\end{equation}
The assumption for single-shot dark-field imaging is that the shift in the centre position is small compared to the width of the curve, i.e. $\Delta \bar{x}_{ref}^2 << \sigma_r^2$, which is generally valid away from material boundaries where the refraction is strongest. The above conditions allows for a simplification in \cref{eq:sampIC_peak} where $\exp\left [ -\frac{\Delta \bar{x}_{ref}^2}{2(\sigma_r^2+\sigma_o^2)} \right ] \approx 1$. The following step involves taking the ratio between \cref{eq:refIC_peak} and \cref{eq:sampIC_peak} and squaring the result to obtain:
\begin{equation} \label{eq:ss_quotient}
    \frac{I_r^2(\bar{x}=\bar{x}_0)}{I_s^2(\bar{x}=\bar{x}_0)} = \frac{\sigma_r^2+\sigma_o^2}{t^2\sigma_r^2}.
\end{equation}
Rearranging this to solve the dark-field signal gives the equation for dark-field retrieval:
\begin{equation} \label{eq:ss_full}
    \sigma_o^2 = \sigma_r^2\left [t^2\frac{I_r^2(\bar{x}=\bar{x}_0)}{I_s^2(\bar{x}=\bar{x}_0)}-1\right ]
\end{equation} 
and defining $\Omega = \sigma_r^2\frac{I_r^2(\bar{x}=\bar{x}_0)}{I_s^2(\bar{x}=\bar{x}_0)}$ this can be written in more compact form as
\begin{equation} \label{eq:ss_full_beta}
    \sigma_o^2 = t^2\Omega - \sigma_r^2
\end{equation}
with the dark-field signal expressed as a function of the width of the reference illumination curves, $\sigma_r^2$, the transmission signal, $t$, and the change in peak illumination curve measurement, quantified as $\Omega$. 

Solving \cref{eq:ss_full_beta} to obtain a dark-field image with a single exposure of the sample still requires knowledge of $t$. An initial approach is that of a phase object, where transmission can be assumed to be unity. For samples showing non-negligible attenuation, simply substituting $t=1$ \cref{eq:ss_full_beta} results in a signal that is unsuitable for tomography because it cannot be expressed as integral as the X-ray path through the sample. By imposing a homogeneous-material approximation it is however possible to establish a relationship between $t$ and $\sigma_0^2$ and constrain the solution of \cref{eq:ss_full_beta}. In the following we show how this provides a signal proportional to the projected sample thickness and is thus compatible with tomographic imaging \cite{endrizzi2017x,doherty2023edge} through standard algorithms such as filtered back projection.  

The relationship between the dark field and attenuation signal is established by defining a new parameter: 
\begin{equation} \label{eq:gamma}
    \gamma = \frac{-ln(t)}{\sigma_o^2}
\end{equation}  
which is a constant for a given sample, with units of \textmu m$^{-2}$, or the inverse of dark-field imaging units. This equation effectively assumes a homogeneous sample i.e. single material. We note that this does not imply a homogeneous microstructure i.e. one without density variations, which would result in a sample with weak scattering. A similar approach was successfully implemented in single-shot phase contrast imaging, where a homogeneous sample allows a linear relationship between the real and imaginary parts of the refractive index, most commonly used with Paganin retrieval \cite{paganin2002simultaneous} \cite{beltran20102d} in free-space propagation, and later translated to other phase-contrast imaging set-ups \cite{burvall2011phase} \cite{diemoz2015single} \cite{zamir2017edge}. 

The second assumption underlying the proposed approach is that transmission is relatively high. This allows for the second-order Taylor expansion in \cref{eq:gamma} to be a good approximation. Replacing $t$ with $1-a$ above gives the following expression after the Taylor expansion around $-ln(1-a)$
\begin{equation}
    -ln(1-a) = a+\frac{a^2}{2}
\end{equation}
where $a$ is the fraction of the beam absorbed by the sample. Note that through the homogeneous material approximation, this is also related to the dark-field signal. Substituting this into \cref{eq:gamma} and rearranging gives the following:
\begin{equation} \label{eq:gamma_linear}
    \sigma_{o}^2 = \frac{1}{\gamma} \left(a+\frac{a^2}{2}\right).
\end{equation}
Substituting this and $t=1-a$ into \cref{eq:ss_full_beta} gives the following polynomial
\begin{equation} \label{eq:paganinSS}
    0 = \left (\gamma\Omega-\frac{1}{2}\right)a^2-\left(2\gamma\Omega+1\right)a+\left (\gamma\Omega-\gamma\sigma_r^2 \right)
\end{equation}
which is a quadratic whose roots can be found as
\begin{figure}[t]
    \centering
    \includegraphics[width=0.5\columnwidth]{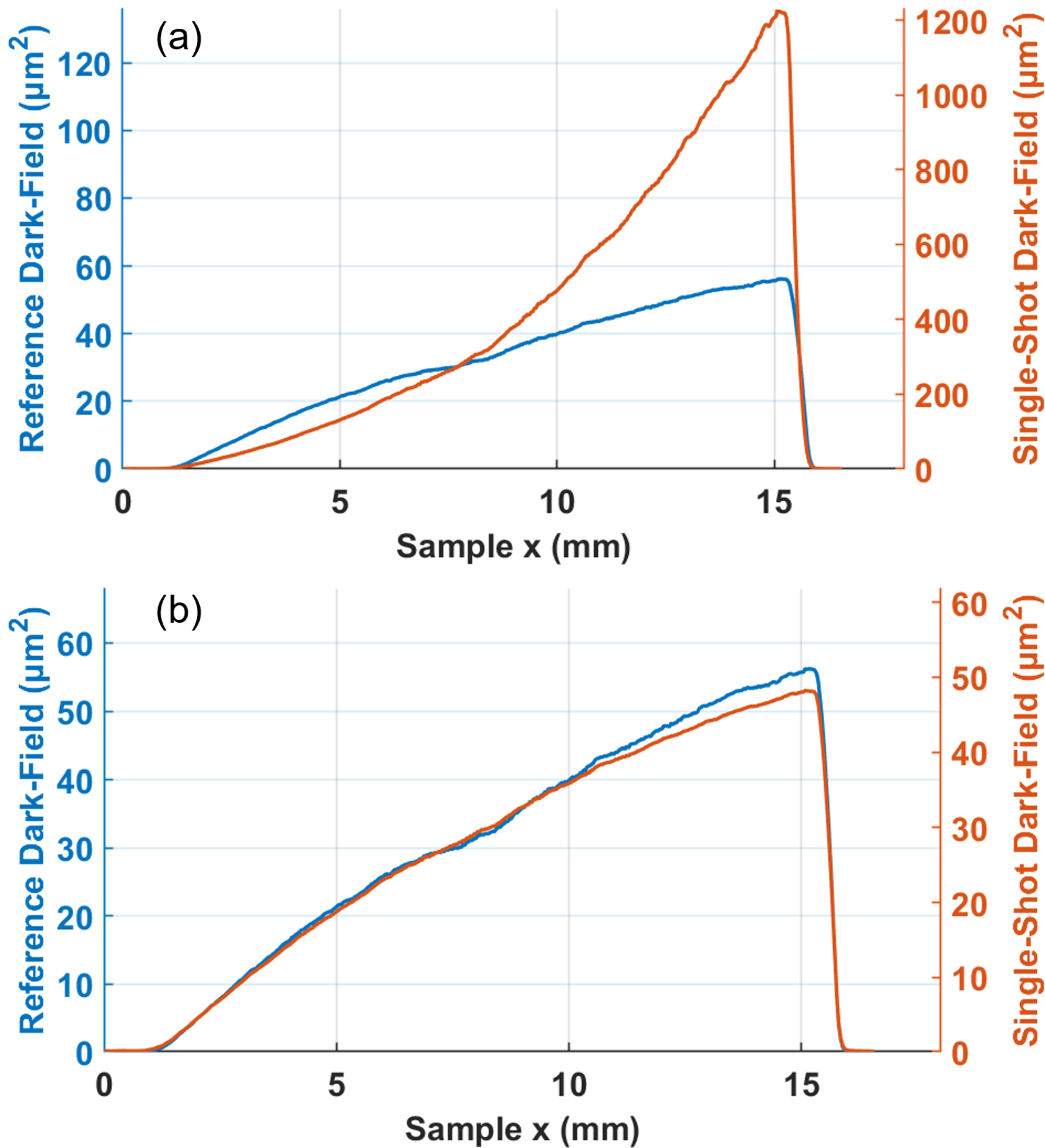}
    \caption{Profiles from paper wedge phantoms. In (a), the single-shot dark-field signal is compared to the reference through \cref{eq:ss_full_beta} with $t=1$. A large discrepancy is found and the signal is overestimated due to the non-negligible contribution from attenuation that is not accounted for. In (b) the single-shot dark-field signal through \cref{eq:gamma_linear} with $\gamma = 0.02$ \textmu m$^{-2}$ is compared to the reference. The reference dark-field signal is retrieved from a data set with full illumination curve sampling with an established phase retrieval algorithm.}
    \label{fig:wedges_linearised}
\end{figure}
\begin{equation} \label{eq:paganinSS_a_solve}
    a = \frac{\left(2\gamma\Omega+1\right)\pm\sqrt{\left(-2\gamma\Omega-1\right)^2-4\left(\gamma\Omega-\frac{1}{2}\right)\left(\gamma\Omega-\gamma\sigma_r^2\right)}}{2\gamma\Omega-1}
\end{equation}
where it was the solution with the negative square root which was found to give the solution that best matched the expected dark-field signal due to the performance in the limit of $\gamma=0$ \textmu m$^{-2}$ . \cref{eq:gamma_linear} can be applied pixel-wise and used to solve for a single-shot dark-field image. It will approximate the true dark-field signal if (i) transmission is high, (ii) a homogeneous object can be assumed, and (iii) $\gamma$ is correctly estimated. The first point holds up to roughly 50\% transmission (See Supplemental Material), which is a substantial extension (i.e. higher attenuation) if compared to the requirement for a pure phase object, needed if \cref{eq:ss_full_beta} is used to retrieve from a single exposure.  The second and third requirements are more difficult to achieve with non-trivial samples. In practice, the single-shot retrieval is unlikely to yield a quantitative dark-field retrieval across the full image, but a qualitative tomographic reconstruction from a linear single-shot signal showing mixed attenuation and dark-field contrast is achievable with this approach.

We first tested this approach on wedges of uniform material, for which a signal that grows linearly with sample thickness is expected. This was found to be the case when using a correct estimate for $\gamma$, in \cref{fig:wedges_linearised} we show results for a paper wedge. We refer the reader to the Supplemental Material for extended results including both weakly and strongly absorbing materials.

\begin{figure*}[t]
    \centering
    \includegraphics[width=\linewidth]{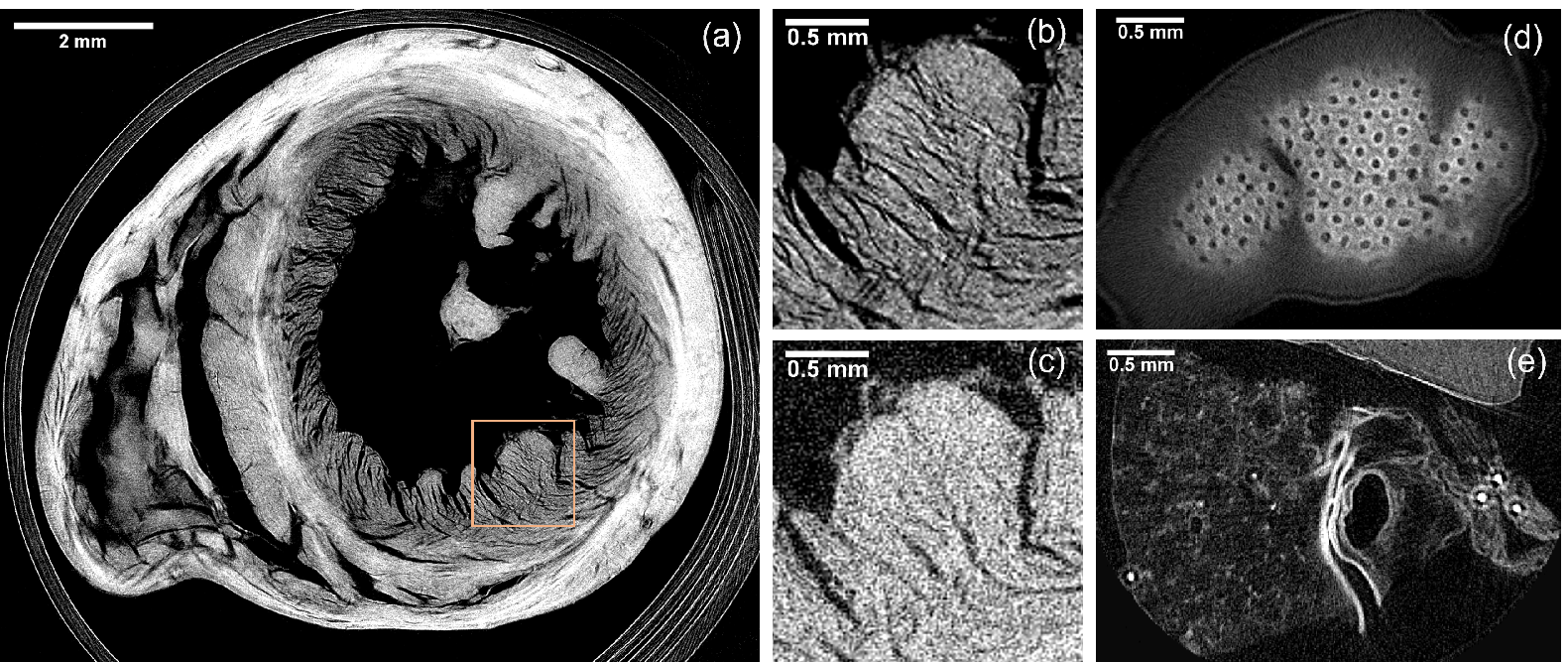}
    \caption{X-ray single-shot dark-field tomography examples of biological tissues. (a) shows a rat heart slice retrieved with \cref{eq:gamma_linear} and $\gamma = 0.0068$ \textmu m$^{-2}$, where the different orientations of fibres give contrast between layers within the heart wall. A section of this slice is shown in (b) indicated by the orange box. This is compared to the same section for a slice with attenuation contrast in (c) from a conventional retrieval, where much poorer contrast as well as a lack of details is observed when the image is built without dark-field sensitivity. In (d) a chicken bone tomography slice with $\gamma = 0.041$ \textmu m$^{-2}$ where the edges of the pores are sharper due to the edge enhancement of the dark-field signal. In (e), a sample of decellularized lung scaffold repopulated with pancreatic islets in the alveolar space has been reconstructed with $\gamma = 0.005$ \textmu m$^{-2}$. Cell clusters are visible (mainly from attenuation contrast) as well as changes in porosity across the scaffold (mainly from dark-field contrast). Boundaries in the decellularized lung scaffold appear bright, which is due to stronger scattering in those regions.}
    \label{fig:tomography}
\end{figure*}

\section*{Single-Shot Dark-Field Tomography}

We demonstrate the application of single-shot dark-field imaging to tomography on examples of biological tissues: a rat heart, a chicken bone, and vascularised endocrine pancreatic tissue. Datasets were acquired with 8 dithering steps, and 1200 angular views through 360$^\circ$, with each frame taken with 1.2 seconds exposure at the top of the illumination curves. The heart was obtained from the UCL Biological Services Unit, from rats euthanised for organ harvesting via Schedule 1 methods, and critically point dried as per Savvidis et al. \cite{savvidis2022monitoring}. The vascularized endocrine pancreas tissue constructs were generated as per the protocol described by Citro et al. \cite{citro2019biofabrication} \cite{citro2023directed}, with pancreatic islets embedded in the alveoli of the decellularized lung tissue. The bone was acquired through commercially available chicken legs and air-dried. The single-shot dark-field projections were retrieved from this data using \cref{eq:gamma_linear} and reconstructed with the standard filtered-back projection algorithm. Estimates for $\gamma$ for each sample were acquired from separate fully sampled datasets, which provided averages of $t$ and $\sigma_o^2$.

All samples were successfully reconstructed with no obvious cupping artefacts are seen from poor signal linearity (see \cref{fig:tomography}). 
\cref{fig:tomography}(a) shows an axial slice through the middle of the rat heart sample. Dark-field contrast highlights the different orientations of muscle fibre in different layers within the heart wall. The region highlighted in orange is expanded in \cref{fig:tomography}(b) and the corresponding area with attenuation contrast in shown in panel (c). The single-shot dark-field tomography image (\cref{fig:tomography}(b)) appears sharper and richer in detail. Further comparison between single-shot dark-field and attenuation contrast tomography for this sample is given in the Supplemental Material. The chicken bone sample is shown in \cref{fig:tomography}(d), with a highlighted porous structure thanks to dark-field contrast sensitivity. The pancreatic construct is shown in \cref{fig:tomography}(e) where a bright layer on the inner surface due to a strong scattering originating there. The bright spots are round-shaped pancreatic islets, and from a full illumination curve planar image, were found to show high attenuation signal. Simultaneous sensitivity to both contrast channels enables concurrent visualisation of the islet and the subtle changes in porosity throughout the decellularised scaffold. We found the tomographic images to have high signal-to-noise ratios with features visible from both contrast channels, although distinguishing the origin of the intensity (i.e. whether it comes from high attenuation or dark field) might not always be possible, it can often be estimated with some sample prior knowledge. 

\section*{Conclusions}

We presented a method to obtain X-ray dark-field images with a single exposure, without the need for a high spatial or temporal coherence or a high-resolution detector. We experimentally demonstrated the approach on a laboratory-based edge illumination imaging system, where the use of two absorbing masks allows sensitivity to dark-field contrast in a single shot through the removal of scattered X-rays. We derived an equation to retrieve the dark-field signal, which can be found in a single measurement on pure-phase objects. This was then extended to work with more attenuating samples, down to approximately 50\% transmission. Our approach relies on a homogeneous material approximation, that enables linking the attenuation and dark-field signals within a sample and thus solving for two unknowns with only one measurement. In practice this results in an image with mixed dark-field and attenuation contrast, with features from both channels being retained. We demonstrated the methodology using edge illumination, however, with some modifications to the model, a similar approach could be applied to other dark-field imaging techniques. One key benefit is the reduction of the problem dimensionality by one, which becomes practically relevant for data-intensive applications such as tomographic imaging. We have shown results for single-shot X-ray dark-field tomography on three biological tissue samples, where the power of dark-field imaging to highlight sub-resolution features was retained with fast and dose-efficient data acquisition. We believe this is a powerful method for obtaining high-quality images whilst, at the same time, it removes the necessity of acquiring multi-dimensional datasets for the extraction of dark-field contrast.

\begin{acknowledgments}
AO is supported by the Royal Academy of Engineering under their Chairs in Emerging Technologies scheme (CiET1819/2/78). This work was supported by the National Research Facility for Lab X-ray CT (NXCT) through EPSRC grants EP/T02593X/1 and EP/V035932/1. This work was supported by the Wellcome Trust 221367/Z/20/Z. This work was supported by EPSRC grant EP/T005408/1 and the Royal Society through IEC\textbackslash R2\textbackslash 192116.
\end{acknowledgments}

\bibliography{ssdf_references.bib}

\end{document}


\preprint{APS/123-QED}

\title{Supplemental Material:\\
Single-shot X-ray Dark-field Tomography}

\author{Adam Doherty}
\email{adam.doherty@ucl.ac.uk}
\author{Ian Buchanan}
\author{Alberto Astolfo}
\author{Savvas Savvidis}
\affiliation{Department of Medical Physics and Biomedical Engineering, University College London, London, WC1E 6BT, UK}
\author{Mattia F. M. Gerli}
\affiliation{UCL Division of Surgery and Interventional Science, Royal Free Hospital, NW3 2PF, London, UK}
\affiliation{Stem Cell and Regenerative Medicine Section, Great Ormond Street Institute of Child Health, University College London, London, WC1N 1EH, UK}
\author{Antonio Citro}
\affiliation{San Raffaele Diabetes Research Institute, IRCCS San Raffaele Scientific Institute, Milan, Italy}
\author{Alessandro Olivo}
\author{Marco Endrizzi}
\affiliation{Department of Medical Physics and Biomedical Engineering, University College London, London, WC1E 6BT, UK}

\date{17th August 2023}

\maketitle

\section{Single-shot dark-field signal linearity as a function of sample's projected thickness}

The general case solution of Eq. 7 in the main text requires knowledge of $t$, but assuming a pure phase object or $t=1$, produces a signal that is mixed dark-field and attenuation, but this is non-linear with X-ray path length. This condition means that the single-shot retrieval is not compatible with tomographic reconstruction unless the attenuation could be assumed to be very low. As such a linearisation approach was developed by assuming a homogeneous sample and introducing a material parameter, $\gamma$, that linked the dark-field and attenuation intensities. This will be shown here to improve the linearity of the retrieval, enabling single-shot tomography at higher levels of X-ray attenuation.
\begin{figure*}[b!]
    \centering
    \includegraphics[width=\linewidth]{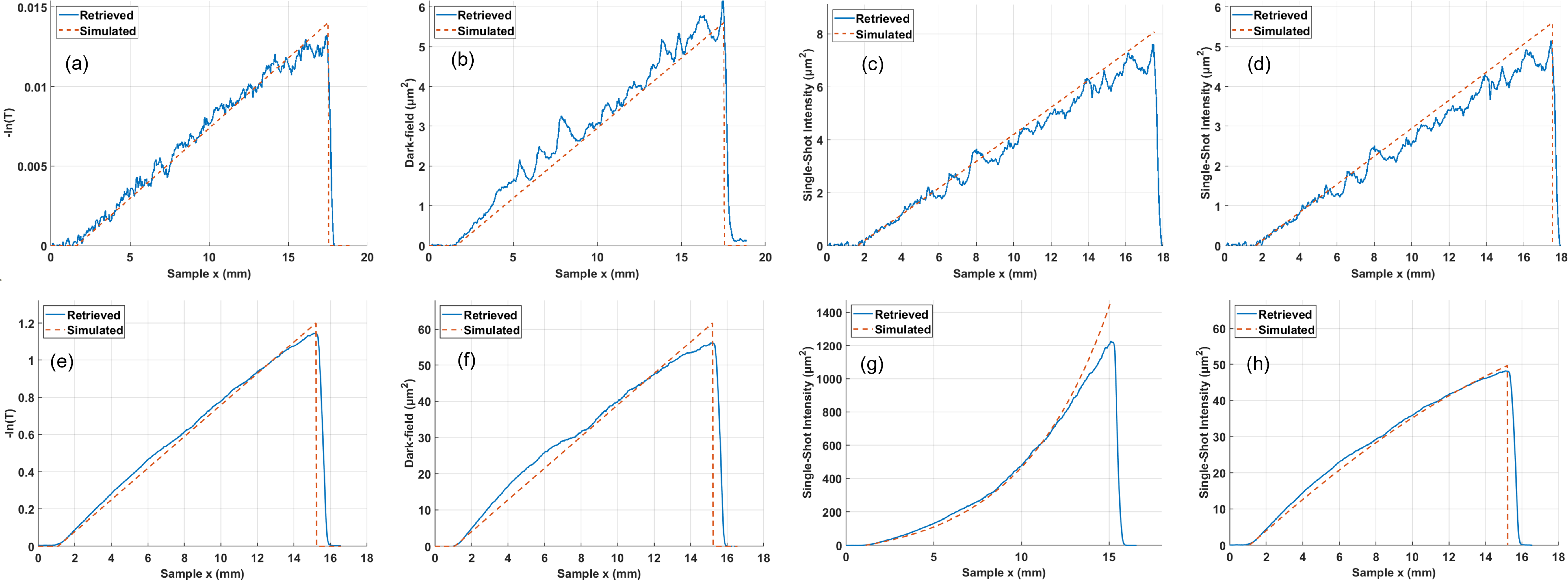}
    \caption{Line profiles extracted from retrieved images of wedges made of polystyrene (top row) and paper (bottom row). Panels (a)+(e) show attenuation extracted from a full illumination curve sampling dataset. Panels (b)+(f) show dark-field contrast extracted from a full illumination curve sampling dataset. Panels (c)+(g) show the single-shot dark-field contrast acquired through Eq. (7) with $t=1$, and panels (d)+(h) show the linearised single-shot dark-field contrast acquired through Eq. (10) with $\gamma = 0.0025$ \textmu m$^{-2}$ for the polystyrene wedge and $\gamma = 0.02$ \textmu m$^{-2}$ for the paper wedge. The phase object approximation $t=1$ works well only for polystyrene, whereas it results in a strongly non-linear signal for a material that attenuates the X-ray beam as well as introduces scattering. An appropriate choice of $\gamma$ allows for the single-shot dark-field signal to be linear with the projected sample's thickness, and hence suitable for tomography application through standard reconstruction algorithms.}
    \label{fig:wedges}
\end{figure*}
We show here experimental images for wedges of two materials that were imaged as test samples for single-shot dark-field imaging. The wedges were made of polystyrene and paper. Data were collected with the complete illumination curve sampling, acquiring 9 illumination curve points and 16 dithering steps, with 1.2 seconds per frame. This dataset also represents the ground truth, against which the single-shot method approach is subsequently bench marked. The single-shot retrieval used the peak illumination curve measurements that were subsampled from this dataset. Experimental data (solid blue lines) are plotted along with results from numerical simulations using the convolutional illumination curve model with attenuation and scattering values equivalent to those measured experimentally (dashed red lines).

The line profiles extracted from the images retrieved from a fully sampled dataset show that attenuation and dark-field signals are approximately proportional to sample thickness as expected. For single-shot dark-field retrieval using Eq. (7) with $t=1$, the intensities are higher than the pure dark-field signal due to the additional contribution from the transmission signal that is not accounted for. This discrepancy grows for samples exhibiting stronger attenuation. The polystyrene wedge can be well approximated as a pure phase object, it shows little attenuation (\cref{fig:wedges}(a)), and hence solving Eq. (7) with $t=1$ provides a reliable result dark-field signal. The introduction of $\gamma$ and solving through Eq. 10 makes little difference. For the case of paper wedge, which is more strongly attenuating, the single-shot retrieval with Eq. 7 and $t=1$ results in a large deviation from linearity which would not be suitable for tomography (\cref{fig:wedges}(g)). With a correct choice of $\gamma$, however, the single-shot retrieval in Eq. (10) vastly improves the linearity of the signal (\cref{fig:wedges}(h)). A slight deviation from linearity remains visible in all plots extracted from the paper wedge dataset, where the signal increases by a smaller proportion as the sample gets thicker. This effect is observed in both the experimental data and simulated data. This is due to the progressively less accurate assumption of low attenuation, breaking the validity of the Taylor expansion and the accuracy of the linearisation approach (See Section \ref{sec:linear_acc}).

\section{Accuracy of single-shot dark-field imaging retrieval as a function of sample transmission}\label{sec:linear_acc}

\begin{figure}[b]
    \centering
    \includegraphics[width=0.6\columnwidth]{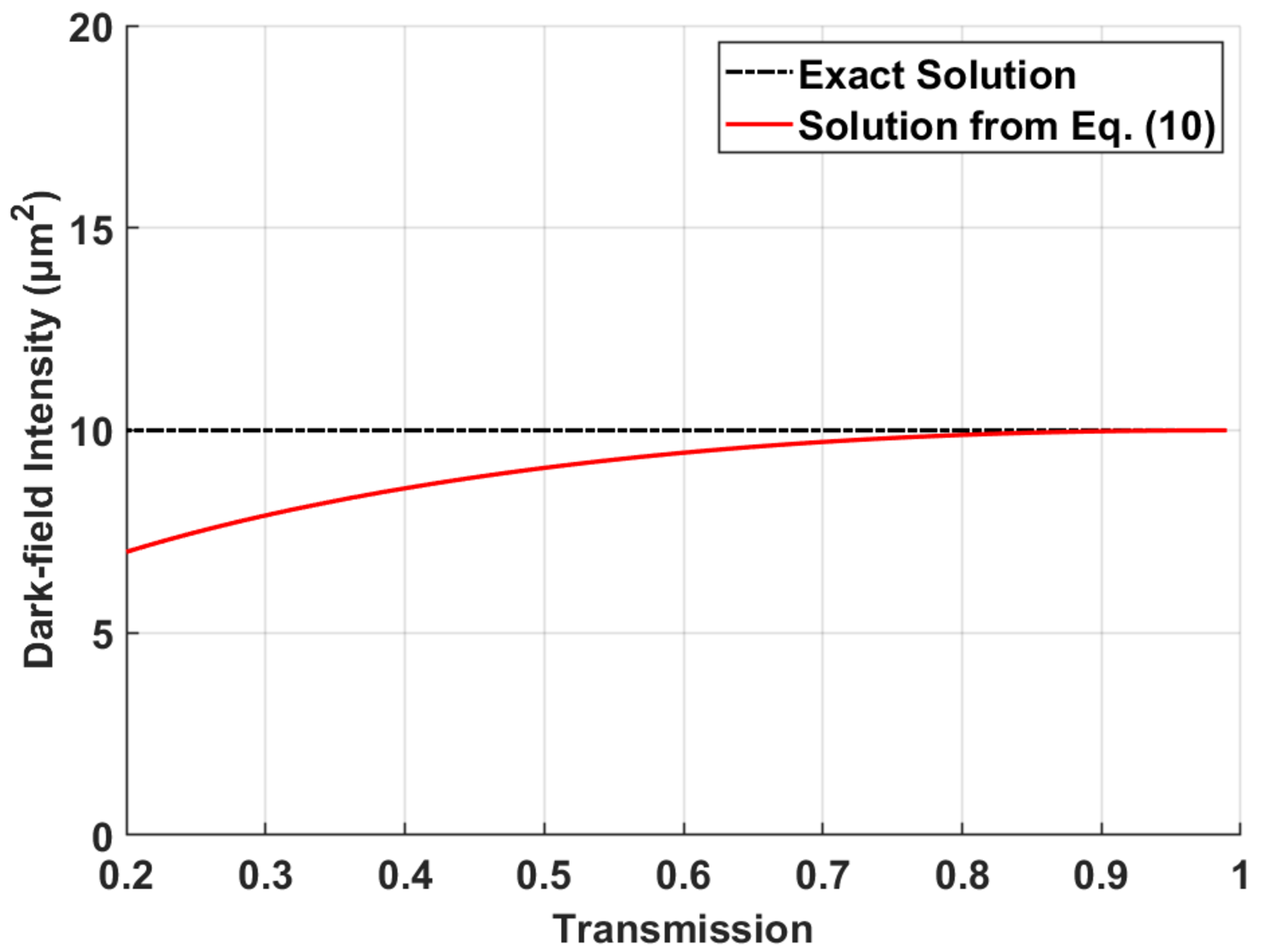}
    \caption{The single-shot signal recovered through Eq. (10) approximates well the exact solution for high transmission with 1\% deviation at $t=0.81$ and 10\% deviation at $t=0.48$.}
    \label{fig:ss_paganin_comp_sim}
\end{figure}
Eq. 10 relies on a Taylor expansion approximation which inevitably becomes less reliable as the transmission of the sample decreases. In practice, however, we have found this approach to be robust in a variety of experiments. We report on a numerical investigation aimed at establishing validity boundaries for this approach, as a function of the sample transmission. The numerical simulation is based on the Gaussian convolution model. Gaussian functions were convolved with scattering functions to model the dark-field signal and reduced in amplitude to model decreasing X-ray transmission. The peak of the resulting Gaussians were then used to quantify $\Omega$, which were then input into Eq. (10). The simulation allowed for a known value of $\gamma$ to be incorporated into the two linearised retrievals. The output from these simulations is shown in \cref{fig:ss_paganin_comp_sim} where the retrieved single-shot dark-field signal is compared against a constant known value for a range of transmission signals.

Eq. (10) is a reliable approximation at high transmission: above $t=0.8$ the error is below 1.1\%. The error reaches 9\% at $t=0.5$, and 30\% at $t=0.2$. The error progressively increases with lower transmission, which in practical terms means that the signal is not linear with sample thickness for high attenuation. As a practical guideline we take 50\% transmission as the limit for the validity of this approach, beyond this the error increases above 10\%. 

\section{Single-shot dark-field compared to attenuation in tomography}

\begin{figure}[b]
    \centering
    \includegraphics[width=\columnwidth, angle = 0]{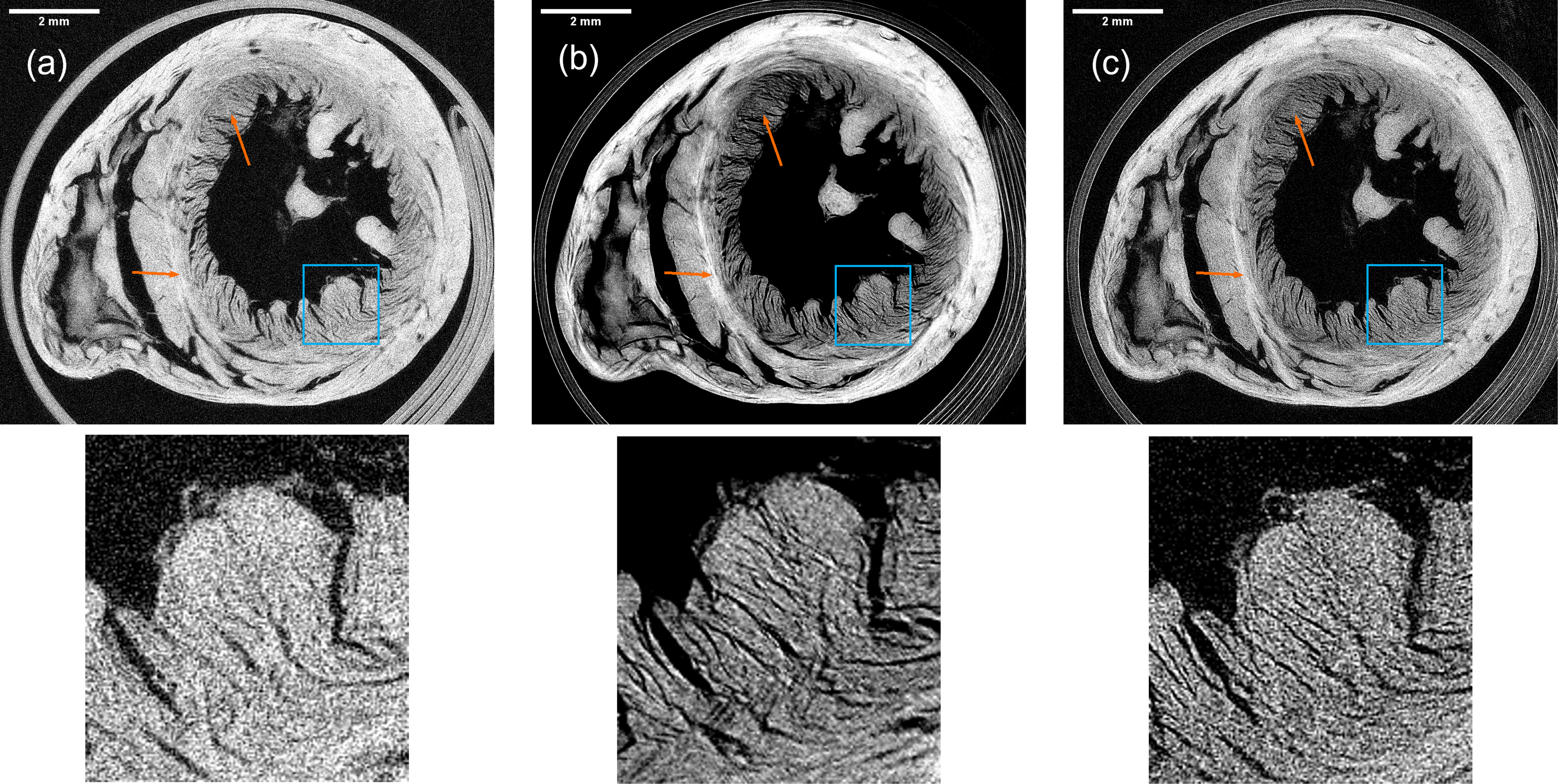}
    \caption{Slices of a rat heart with (a) attenuation contrast, (b) single-shot dark-field contrast with matched exposure and (c) single-shot dark-field contrast with lower exposure. Highlighted structures include a bright layer in the ventricular wall (left arrow) due to the dark-field sensitivity to fibre orientation, and darker intensity seen in the endocardium (right arrow), both only seen with dark-field contrast. The zoom-in around the blue box shows how features within the inner myocardium are visible in the single-shot dark-field reconstructions, which are clear even within the faster scan in the single-shot dark-field reconstruction.}
    \label{fig:heart_comparison}
\end{figure}
Single-shot dark-field tomography combines attenuation and dark-field contrasts in a single image. This means that strictly speaking, the approach is no longer quantitative. However, we have observed that this approach provides images with high signal-to-noise ratios and rich with detail. We report here a comparison between attenuation-contrast tomography and single-shot dark-field tomography to highlight the similarities and differences between these approaches. The sample was scanned with 1200 angular projections and 8 dithering steps. The attenuation image was retrieved by taking 7 illumination curve measurements across the aperture period, with each exposure lasting 1.2 seconds. The single-shot dark-field reconstructions were obtained with a single exposure at the peak of the illumination curves. This was carried out with 1.2 seconds exposure for a faster scan than the fully sampled dataset used for the attenuation image, as well as 8.4 seconds for matched scan time to the fully sampled dataset. Note that the latter has the same dose and exposure time as the fully sampled illumination curve dataset, whereas the former carries a 7-fold speed-up or dose reduction. As highlighted in  \cref{fig:heart_comparison}, single-shot dark-field images offer enhanced contrast between muscle layers as well as richer details when compared to the more conventional attenuation-contrast.